\documentclass[sigconf]{acmart}

\usepackage{booktabs} 
\usepackage{subcaption}

\AtBeginDocument{%
  \providecommand\BibTeX{{%
    \normalfont B\kern-0.5em{\scshape i\kern-0.25em b}\kern-0.8em\TeX}}}

\setcopyright{acmcopyright}
\copyrightyear{2018}
\acmYear{2018}
\acmDOI{10.1145/1122445.1122456}

\acmConference[Woodstock '18]{Woodstock '18: ACM Symposium on Neural
 Gaze Detection}{June 03--05, 2018}{Woodstock, NY}
\acmBooktitle{Woodstock '18: ACM Symposium on Neural Gaze Detection,
 June 03--05, 2018, Woodstock, NY}
\acmPrice{15.00}
\acmISBN{978-1-4503-9999-9/18/06}

\setcopyright{none}
\copyrightyear{}
\acmYear{}
\acmDOI{}
\acmConference[]{}{}{}
\acmBooktitle{}
\acmPrice{}
\acmISBN{}

\makeatletter
\renewcommand\@formatdoi[1]{\ignorespaces}
\makeatother

\newcommand\blfootnote[1]{%
  \begingroup
  \renewcommand\thefootnote{}\footnote{#1}%
  \addtocounter{footnote}{-1}%
  \endgroup
}

\begin{document}

\settopmatter{printacmref=false} 
\renewcommand\footnotetextcopyrightpermission[1]{} 
\pagestyle{plain} 

\title{Rendering Point Clouds with Compute Shaders}

\author{Markus Sch{\"u}tz}
\affiliation{%
  \institution{TU Wien}
  \city{Vienna}
  \country{Austria}}
\email{mschuetz@cg.tuwien.ac.at}

\author{Michael Wimmer}
\affiliation{%
  \institution{TU Wien}
  \city{Vienna}
  \country{Austria}}
\email{wimmer@cg.tuwien.ac.at}

\renewcommand{\shortauthors}{Trovato and Tobin, et al.}

\begin{abstract}

We propose a compute shader based point cloud rasterizer with up to 10 times higher performance than classic point-based rendering with the GL\_POINT primitive. In addition to that, our rasterizer offers 5 byte depth-buffer precision with uniform or customizable distribution, and we show that it is possible to implement a high-quality splatting method that blends together overlapping fragments while still maintaining higher frame-rates than the traditional approach.

\end{abstract}

%
%



\begin{teaserfigure}
	\begin{subfigure}[b]{0.246\textwidth}
		\centering
		\includegraphics[height=2.4cm]{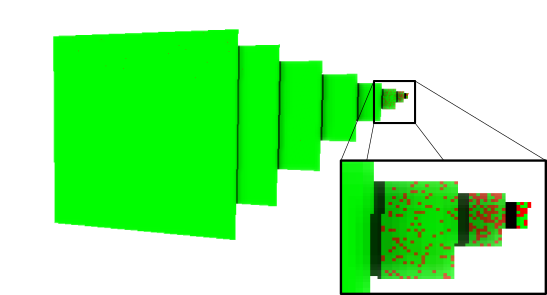}
		\caption{Regular OpenGL z-fighting}
		\label{fig:gull}
	\end{subfigure}
	~ 
	\begin{subfigure}[b]{0.246\textwidth}
		\centering
		\includegraphics[height=2.4cm]{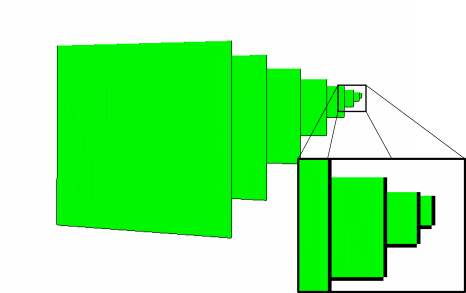}
		\caption{40 bit integer depth buffer}
		\label{fig:tiger}
	\end{subfigure}
	~ 
	\begin{subfigure}[b]{0.246\textwidth}
		\centering
		\includegraphics[height=2.4cm]{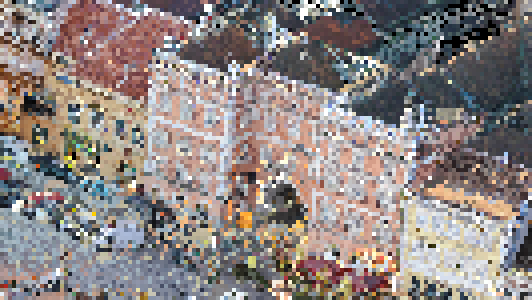}
		\caption{Regular Rasterization}
		\label{fig:mouse}
	\end{subfigure}
	~
	\begin{subfigure}[b]{0.246\textwidth}
		\centering
		\includegraphics[height=2.4cm]{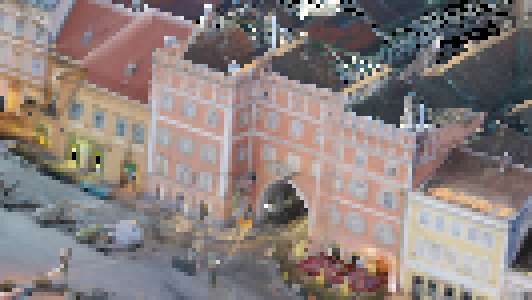}
		\caption{High-Quality Splatting}
		\label{fig:mouse}
	\end{subfigure}
	\caption{Point-based rendering via compute shaders. (b) Higher depth-precision. (c) Up to ten times faster than OpenGL rasterizer with regular quality. (d) Two to three times faster with high-quality splatting.}\label{fig:teaser}
\end{teaserfigure}

\maketitle

\blfootnote{Note: This is a non-peer-reviewed poster abstract that is currently being submitted to a peer-reviewed venue. Source code and implementation details are available at https://github.com/m-schuetz/compute\_rasterizer}

\section{Introduction}

Traditionally, point clouds in OpenGL are rendered with the 
glDrawArrays(GL\_POINT, ...) command, which passes point primitives through the OpenGL rendering pipeline. While many parts are programmable nowadays, others remain fixed. An appealing quality of GPGPU is that it gives developers the possibility to write their own rendering pipeline \cite{Kenzel:2018}. G{\"u}nther et al. \cite{Gnther2013AGP} proposed an OpenCL-based point-cloud renderer back in 2013, but were limited to 32-bit atomic operations at the time. Instead of using atomicMin, they implemented a busy loop with an early-out optimization to achieve major performance improvements over OpenGL.

In many cases, especially triangle rendering, the regular rendering pipeline remains faster than GPGPU rasterizers. However, GPGPU allows implementing features that may not be possible in the regular pipeline, such as improved depth buffers. The classic OpenGL projection matrices map nearby depth values over most of the available depth-buffer range, while leaving only little precision to farther parts of the scene. To make thing worse, vertex transformations and resulting depth values are processed with floating-point numbers, which have a higher precision close to zero. A well-known trick to improve the precision is to reverse the depth buffer and map the near-clip plane to 1, and the far-clip plane to 0, so that distant depth values are sampled at a higher precision. Further information can be found in NVIDIA's "Depth Precision Visualized" article \cite{Reed18}. Nonetheless, depth precision is inevitabely lost during the vertex transformation, and the subsequent storage of the result in a single-precision floating-point vector. Our approach improves depth precision by computing the depth with double precision, and storing the result in a 40-bit integer buffer.


\section{Method}

We developed two approaches to draw point clouds with compute shaders instead of the classic rendering pipeline. The first method uses atomicMin to write the closest point into a custom framebuffer. The second method implements high-quality surface splatting \cite{Botsch2005} based on the first method. 

\subsection{Rasterization via AtomicMin}

This approach encodes color and depth into a single 64-bit integer and uses atomicMin to write the closest fragment into a shader storage buffer (SSBO) that acts as our framebuffer. RGB values are stored in the least significant, and the depth in the most significant bits. Due to this, atomicMin primarily takes the depth into account when it writes the value into the framebuffer, except when two fragments have exactly the same depth. In the latter case, the fragment with the smaller color value is picked.

Our approach gives developers full control over a 40-bit integer depth value. Unlike the traditional pipeline, this value is not clamped, and it provides a uniform or customizable and easily predictable precision over the whole range. 40-bits are sufficient to represent 1 trillion different values. Assuming millimeter precision, we end up with 1 trillion mm = 1 billion m = 1 million km, which means we can represent the depth value of any object on earth and as far away as the moon in millimeter precision. To obtain millimeter precision in a scene that is represented in meters, we compute the depth in double precision, multiply it by 1000, and store the integer part in an int64\_t type value. It is also possible to split the full range of depth into sub-ranges with different precision, if higher precision near the camera is required without sacrificing view distance. A progression with, for example, half the precision at double the distance may be a reasonable choice, but functions such as log and pow do not work on double values at this time. Instead, developers can manually map depth ranges to different precisions, for example, [0m, 10m] to nanometers, [10m, 10km] to micrometers and [10km, 10.000km] to millimeters. Each of these ranges occupy at most 10 billion integer values for a total of 30 billion out of 1 trillion available values.

The 64-bit integer depth and the rgb colors are then encoded into a single 64-bit integer. The depth value is shifted 24 bits to the left, reducing its available range to 40 bits, and the color value is stored in the rightmost 24 bits. AtomicMin is then used to write this 64 bit integer into the SSBO. The atomic min operation stores new fragments only if the encoded depth value is smaller than previously written fragments.

In the second step, another compute shader that runs on each pixel reads the values from our custom framebuffer and stores the color values in an actual OpenGL texture. The shader also clears our framebuffer at the end by setting each value to 0xffffffffff000000. The first five bytes are the depth component which are reset to the maximum value, and the last three bytes are the RGB component which act as the background color. If set to zero, the background will be black. 

\begin{table}
	\setlength{\abovecaptionskip}{0pt}
	\caption{Rendering times for Heidentor (26M points), Retz (145M points on 2080 TI, 120M on 1060 GTX) and Morro Bay (117M points).
	}
	\label{tab:benchmarks}
	\begin{tabular}{llrrr}
		\toprule
		Model       & GPU         & AtomicMin & Splatting     & GL\_POINT  \\		
		\midrule
		Heidentor   & 2080 TI     &  1.64 ms  &   3.37 ms     &   5.71 ms   \\
		            & 1060 GTX    &  4.88 ms  &  11.78 ms     &  13.60 ms   \\
		Retz        & 2080 TI     &  6.41 ms  &  12.95 ms     &  34.04 ms   \\
		            & 1060 GTX    & 14.32 ms  &  31.76 ms     &  58.82 ms   \\
		Morro Bay   & 2080 TI     & 5.87 ms   &  15.48 ms     &  60.26 ms   \\
		\bottomrule
	\end{tabular}
\end{table}

\subsection{High-Quality Splatting}

The second approach is an implementation of \emph{High-Quality Surface Splatting on Today's GPUs} \cite{Botsch2005} with compute shaders. It achieves anti-aliasing by computing an average of the closest fragments within a pixel. Many of the points in a pixel are samples of the same front-most surface and therefore all of them should contribute to the pixel. In basic rendering approaches, however, only the closest fragment is drawn. 

Our compute-based version works as follows. The first pass creates a depth-buffer using the atomicMin approach from the previous section. The second pass sums up the red, green and blue values of all fragments who's linear depth values are at most 1\% larger than the previously computed closest depth in that fragment's pixel. We use a percentage because it makes this method work at arbitrary distances. Each fragment that passes the depth-test also increments the fragment counter for that pixel. In the third and last pass, the final color value of a pixel is computed by dividing the sum of fragment colors by the number of fragments. The result is an image where each pixel contains the average of overlapping points within a certain depth range, rather than only the closest point.

\section{Performance}

Table~\ref{tab:benchmarks} compares rendering times of our two compute based methods against the traditional GL\_POINT method. For example, Retz on a 2080 TI renders 5.3 times faster, and Morro bay renders 10 times faster than GL\_POINTs. The high-quality splatting method renders 2.6 times and 3.9 times faster for the respective data sets. We would like to note that the results vary greatly depending on the order of points and the selected viewpoint, and shuffling points reduces the efficiency of our compute based method. More detailed benchmarking will be part of future work.

\section{Conclusions and Future Work}

We have shown that in the context of point clouds, compute shaders are not only a viable but possibly advantageous alternative to the traditional OpenGL rendering pipeline, with speed-ups of up to 10 times. However, at this time all work was done and evaluated on point sizes of one pixel. Initial tests have shown that our current compute shader implementation scales roughly linearly with the number of pixels per point, wheres the OpenGL rasterizer scales better than that. Our approach is therefore ideal for point sizes of 1 pixel, but less suited for sizes larger than 2 pixels.

We believe that compute based point rasterizers will be useful for web-based rendering with the upcoming WebGPU API. On Microsoft Windows, WebGL is translated to DirectX, which does not support sized point sprites. The translation, ANGLE, therefore emulates the GL\_POINT primitive, which results in a significant loss of performance. With WebGPU, developers may be able to benefit from the improved performance and quality of our compute shader method.

\begin{acks}
  The authors would like to thank the \emph{Ludwig Boltzmann Institute for Archaeological Prospection and Virtual Archaeology} for the Heidentor data set, \emph{Riegl} for the data set of the town of Retz,
  and \emph{PG\&e} and \emph{Open Topography} for providing and hosting the Morro Bay data set.

\end{acks}

\bibliographystyle{ACM-Reference-Format}
\bibliography{compute}

\end{document}